\documentclass[a4paper,11pt]{article}
\usepackage{pos}
\usepackage{graphicx} 
\usepackage{slashed}
\usepackage{hyperref}

\def\={\stackrel{\bullet}{=}}

\def\({\left(}
\def\){\right)}
\def\[{\left[}
\def\]{\right]}

\def \be {\begin{equation}}
\def \ee {\end{equation}}
\def \beqa {\begin{eqnarray}}
\def \eeqa {\end{eqnarray}}
\def \beal#1 {\begin{align}#1\end{align}}
\def \bes#1 {\begin{equation}\begin{split}#1\end{split}\end{equation}}

\def\x't{(\boldsymbol{x'},t)}

\def\3tensor#1#2#3#4{#1^{#2\;#4}_{\;\;#3}}

\def\i{\mathrm{i}}

\def\ccr{c_{\boldsymbol{r}}}

\title{Conserved Non-Singlet Charges for Staggered Fermion Hamiltonian in 3+1 Dimensions}

\author*[a]{Tetsuya Onogi}
\author[a]{Tatsuya, Yamaoka}

\affiliation[a]{IDepartment of Physics, The University of Osaka,\\
Toyonaka, Osaka 560--0043, Japan}

\emailAdd{onogi@hetmail.phys.sci.osaka-u.ac.jp}

\abstract{We study conserved charges of the staggered fermion Hamiltonian in
3+1 dimensions.
By decomposing staggered fermions into Majorana components and exploiting
lattice translation symmetries, we construct a set of conserved
non-singlet charges.
We analyze their algebra and show that, although the charges exhibit
nontrivial non-commutativity on the lattice, they generate axial
$SU(2)_L \times SU(2)_R$ transformations for low-energy degrees of freedom
in the continuum limit.
Possible implications for anomalies are discussed.
}

\FullConference{The 42nd International Symposium on Lattice Field Theory (LATTICE2025)\\
2-8 November 2025\\
Tata Institute of Fundamental Research, Mumbai, India\\}

\begin{document}
\maketitle

\section{Introduction}
\label{sec:intro}

Staggered fermions provide an economical lattice formulation of fermions
by reducing the number of fermion doublers while preserving a remnant of
chiral symmetry.
The realization of flavor and chiral symmetries in this framework has
been a longstanding issue, particularly in four dimensions.
In 1+1 dimensional systems, conserved charges associated with staggered
fermions and their physical interpretation have been studied in detail \cite{Chatterjee:2024gje, Yamaoka:2025sdm}.
The Hamiltonian is given as 
\begin{eqnarray}
H = -i \sum_j (c_j^\dagger c_{j+1} + c_j c_{j+1}^\dagger) .
\end{eqnarray}
They found that there exist integer-valued conserved vector and axial charges $Q_V, Q_A$.
Moreover, $[Q_V, Q_A]\neq 0$ holds, which reflects the Nielsen-Ninomiya theorem. They found that 
\begin{enumerate}
\item $Q_V, Q_A$ are part of the Onsager algebra\cite{Onsager:1943jn}.
\item If we impose both symmetries, possible local terms $\delta H$ which you can add to the hamiltonian is strongly restricted, i.e.
a) Only bilinear terms are allowed and b) mass terms are prohibited. 
\end{enumerate}
This observation gives a hint to consider possible symmetric mass generation.  
A natural question is whether we can extend this finding to the staggered fermion in 3+1 dimensions.

In these proceedings, we report our study of conserved charges in the staggered fermion Hamiltonian in $3+1$ dimensions\cite{Onogi:2025xir}.
The staggered fermion Hamiltonian in $3+1$D has been analyzed in detail in 
Refs.~\cite{Susskind:1976jm,Catterall:2025vrx,Li:2024dpq}.
%
%
Using the Stern transformation, we clarify the physical interpretation of these conserved charges and analyze their algebraic structure. Particular attention is given to the interplay between lattice-specific non-commutativity and its possible implications for anomalies in the continuum limit.

\section{Staggered Fermion Hamiltonian and Symmetries}
\label{sec:model}
We consider the staggered fermion Hamiltonian in $3+1$ dimensions~\cite{Kogut:1974ag,Banks:1975gq,Susskind:1976jm,Catterall:2025vrx,Li:2024dpq}.,
\begin{eqnarray}
H := i \sum_{\mathbf{r},i} \eta_i(\mathbf{r})\,c_{\mathbf{r}}^\dagger( \nabla_i c)_{\mathbf{r}} := \frac{i}{2} \sum_{\mathbf{r},i} \eta_i(\mathbf{r})\,c_{\mathbf{r}}^\dagger (c_{\mathbf{r}+\hat{x}_i}-c_{\mathbf{r}-\hat{x}_i}) ,
\label{eq:staggeredH}
\end{eqnarray}
where $\ccr$ 
satisfies the  anticommutation relation
$    \{c_{\boldsymbol{r}},c_{\boldsymbol{r'}^\dagger}\} = \delta_{\boldsymbol{r},\boldsymbol{r'}} $.
Here, $r$ labels lattice sites and  $i=1,2,3$. $\hat{x}_i$ is the unit vector in the $i-$th direction, and $\eta_i(\mathbf{r}):= (-1)^{r_1+\cdots+r_{i-1}}$ are the usual
staggered phase factors.
The Hamiltonian is invariant under lattice translations and possesses
vector $U(1)_V$ symmetry.
Additional symmetries emerge when the Majorana structure is taken into
account.

%
\subsection{Construction of Conserved Charges}
\label{sec:charges}
The conserved charges are constructed in the following steps:

\begin{enumerate}
\item Decompose complex fermions into Majorana fermions $a$ and $b$.
\item Identify lattice translation symmetries acting on the $b$-fermions.
\item Construct conserved charges using translated $b$-fermions.
\item Classify allowed operators using lattice symmetries.
\end{enumerate}
To analyze the symmetry structure, it is convenient to decompose the
complex fermion into two Majorana fermions as 
$c_\mathbf{r}= \frac{1}{2} \left( a_\mathbf{r}+ i b_\mathbf{r}\right) $ 
, the Hamiltonian becomes
\begin{eqnarray}
H = \dfrac{i}{4} \sum_{r,i} \eta_i(\mathbf{r}) ( a_{\mathbf{r}} a_{\mathbf{r}+\hat{x}_i} +b_{\mathbf{r}} b_{\mathbf{r}+\hat{x}_i} )
\label{eq:majorana_H}
\end{eqnarray}
In this representation, lattice translations act differently on the
$a$- and $b$-fermions, which plays a crucial role in constructing
conserved charges.

\noindent
The fundamental conserved $U(1)_V$ charge is defined as
\begin{equation}
Q_0 = \sum_\mathbf{r} c_\mathbf{r}^\dagger c_\mathbf{r} = \frac{i}{2} \sum_\mathbf{r} a_\mathbf{r} b_\mathbf{r}.
\label{eq:Q0}
\end{equation}

To construct additional conserved charges, we define a translation operator $T_{\hat{x}_i}^{(b)}$ that acts only on the Majorana fermions $b_{\boldsymbol{r}}$ along the direction $\hat{x}_i$, such that
\begin{eqnarray}
    T_{\hat{x}_i}^{(b)} a_{\boldsymbol{r}} \left(T_{\hat{x}_i}^{(b)}\right)^{-1} = a_{\boldsymbol{r}}, &&
    T_{\hat{x}_i}^{(b)} b_{\boldsymbol{r}} \left(T_{\hat{x}_i}^{(b)}\right)^{-1} = \zeta_i(\boldsymbol{r}) \, b_{\boldsymbol{r} + \hat{x}_i}  ,
\end{eqnarray}
where $\zeta_i(\boldsymbol{r}) := (-1)^{\sum_{k=i+1}^3 x_k}$~\cite{Catterall:2025vrx}. The Hamiltonian is manifestly invariant under this transformation.

This action can be extended to arbitrary lattice directions $\boldsymbol{\chi} = n_x \hat{x} + n_y \hat{y} + n_z \hat{z}$, where $n_i \in \mathbb{Z}$, through a composite operator
\begin{eqnarray}
    T_{\boldsymbol{\chi}}^{(b)} a_{\boldsymbol{r}} \left(T_{\boldsymbol{\chi}}^{(b)}\right)^{-1} = a_{\boldsymbol{r}} , &&
    T_{\boldsymbol{\chi}}^{(b)} b_{\boldsymbol{r}} \left(T_{\boldsymbol{\chi}}^{(b)}\right)^{-1} =
    \left(\zeta_y(\boldsymbol{r})\right)^{n_y} \left(\zeta_x(\boldsymbol{r})\right)^{n_x} \, b_{\boldsymbol{r} + \boldsymbol{\chi}}  .
\end{eqnarray}

Using the translation operators, we define new conserved charge operators by conjugating $Q_0$,
\begin{eqnarray}
    \label{eq:Q-chi-operator}
    Q_{\boldsymbol{\chi}} := T_{\boldsymbol{\chi}}^{(b)} Q_0 \left(T_{\boldsymbol{\chi}}^{(b)}\right)^{-1}
   = \frac{\i}{2} \sum_{\boldsymbol{r} \in \Lambda} \left(\zeta_y(\boldsymbol{r})\right)^{n_y} \left(\zeta_x(\boldsymbol{r})\right)^{n_x} a_{\boldsymbol{r}} b_{\boldsymbol{r} + \boldsymbol{\chi}} .
\end{eqnarray}
Using lattice translations $T_i^{(b)}$ acting on the $b$-fermions, we
define additional conserved charges,
\begin{equation}
Q_i = T_i^{(b)} Q_0 T_i^{(b)\,-1} ,
\qquad i=1,2,3 .
\label{eq:Qi}
\end{equation}
One can show that these charges commute with the Hamiltonian,
\begin{equation}
[H, Q_0] = [H, Q_i] = 0 .
\end{equation}

\subsection{Algebra of Conserved Charges}
\label{sec:algebra}
Next, we study the algebra among the conserved charges.
It turns out that the charges on the lattice do not commute :
\begin{eqnarray}
    \label{eq:noncommutativity-Q0-Qxi}
    \left[ Q_0 , Q_{\boldsymbol{\chi}} \right]
   = \frac{\i}{2} \sum_{\boldsymbol{r}} \left(\zeta_y(\boldsymbol{r})\right)^{n_y} \left(\zeta_x(\boldsymbol{r})\right)^{n_x}
    \left( a_{\boldsymbol{r}} a_{\boldsymbol{r} + \boldsymbol{\chi}} - b_{\boldsymbol{r}} b_{\boldsymbol{r} + \boldsymbol{\chi}} \right) 
   \neq 0 .
\end{eqnarray}
%
%
%
While we find that the two generators $Q_0$ and $Q_3$ generates Ons1 algebra\cite{Onogi:2025xir},  it was found by the authors of Ref.~\cite{Aoki:2025vtp, Aoki-lat}  that the conserved charges generate generalized Onsager algebra (Ons3). The algebra strongly constrains symmetry-allowed operators in the Hamiltonian and plays an important role in determining whether the system can acquire a mass gap.

\section{Physical Interpretation of Charges}
\label{sec:interpretation}

To clarify the physical interpretation of the conserved charges, we rewrite the 
system  using a real-space Stern transformation, the staggered fermion degrees
of freedom are mapped to matrix-valued fermions. This transformation was used to give find the physical interpretation for staggered Hamiltonian 1+1 dim~\cite{Yamaoka:2025sdm}.
We denote the lattice points with even integer values in $x, y$ directions as
\begin{eqnarray}
\mathbf{R}_e = (2n_x, 2n_y, n_z), & n_x, n_y, n_z \in \mathbf{Z}
\end{eqnarray}
Introducing the matrices constructed from the gamma matrices,
\begin{eqnarray}
    \label{eq:alphaI-alphaR}
    \alpha_i = \gamma_0 \gamma_i  ,  &\beta = \gamma_0 \ ,   \alpha^r &= \alpha_1^{x} \alpha_2^{y} \alpha_3^{z} \ ,
\end{eqnarray}
which satisfy the algebraic relations
   $ \{ \alpha_i , \alpha_j \} = 2 \delta_{ij} \ ,  \{ \alpha_i , \beta \} = 0 \ ,  \alpha_i^\dagger = \alpha_i $. 
We employ the Weyl representation,
\begin{eqnarray}
    \label{eq:gamma-WeylRep}
    \gamma_0 = \mathbb{I}_{2\times 2} \otimes \sigma_1 
      = 
      \begin{pmatrix}
        0 & \mathbb{I}_{2\times 2} \\
        \mathbb{I}_{2\times 2} & 0
    \end{pmatrix} ,
    & 
        \gamma_i = \sigma_i \otimes \i \sigma_2 
        = 
        \begin{pmatrix}
        0 & \sigma_i \\
        - \sigma_i & 0
    \end{pmatrix}, 
\end{eqnarray}
which leads to
\begin{eqnarray}
    \gamma_5 
    = \mathbb{I}_{2\times 2} \otimes \sigma_3 
    =
     \begin{pmatrix}
        \mathbb{I}_{2\times 2} & 0 \\
        0 & - \mathbb{I}_{2\times 2}
    \end{pmatrix} ,
    &
        \alpha_i = - \sigma_i \otimes \sigma_3 
        = 
        \begin{pmatrix}
        - \sigma_i & 0 \\
        0 & \sigma_i
    \end{pmatrix} .
\end{eqnarray}

We then define a matrix-valued fermion field from the staggered fermions as follows~\cite{Catterall:2025vrx},
\begin{align}
    \label{eq:Matrix-Fermion}
    \Phi ( \boldsymbol{R}_e ) &= N_0 \sum_{\{ \boldsymbol{b} \}} c_{\boldsymbol{R}_e + \boldsymbol{b}} \, \alpha^{b} 
   = \begin{pmatrix}
        \psi_{1+}(\boldsymbol{R}_e) & \psi_{2+}(\boldsymbol{R}_e) & 0 & 0 \\
        0 & 0 & \psi_{3-}(\boldsymbol{R}_e) & \psi_{4-}(\boldsymbol{R}_e)
    \end{pmatrix} \ ,
\end{align}
where $\{\boldsymbol{b}\}$ is a set of $2^3$ vectors with components $b_i = \{0,1\}$, 
$\boldsymbol{R}_e = 2 \boldsymbol{r}$, $N_0$ is a normalization constant, and $\psi_{1,2}$ ($\psi_{3,4}$) are two-component left-handed (right-handed) Weyl fermions.

We now combine the Weyl fermions $\psi_f$ into two four-component Dirac fermions,  
\begin{align}
    \psi_1 (\boldsymbol{R}_e) &= \begin{pmatrix} \psi_{1+}(\boldsymbol{R}_e) \\ \psi_{3-}(\boldsymbol{R}_e) \end{pmatrix} \, , \quad 
    \psi_2 (\boldsymbol{R}_e) = \begin{pmatrix} \psi_{2+}(\boldsymbol{R}_e) \\ \psi_{4-}(\boldsymbol{R}_e) \end{pmatrix} \, ,
\end{align}
which together form a $2 \times 2$ matrix-valued fermion field,
\begin{align}
    \label{eq:2-2-matrix-fermion}
    \Psi (\boldsymbol{R}_e) 
    = \begin{pmatrix} \psi_1(\boldsymbol{R}_e) & \psi_2(\boldsymbol{R}_e) \end{pmatrix} 
    = \begin{pmatrix}
        \psi_{1+} & \psi_{2+} \\
        \psi_{3-} & \psi_{4-}
    \end{pmatrix} \, .
\end{align}
The Hamiltonian~\eqref{eq:staggered-Hamiltonian} can then be written as
\begin{align}
    \label{Hamiltonian-matrix-2-flavor}
    H &= \i \sum_{\boldsymbol{R}_e} \Psi^\dagger(\boldsymbol{R}_e) \left[
        (\alpha_i \otimes \mathbb{I}_{2 \times 2}) \frac{\nabla_i}{2}  
        + (\beta \gamma_5 \otimes \sigma_i^T) \frac{\nabla_i^2}{2} \right] \Psi(\boldsymbol{R}_e) \, ,
\end{align}
where the matrices $\alpha_i$ and $\beta \gamma_5$ act on the spin indices, while $\sigma_i^T$ acts on the flavor indices. The lattice difference operator $\nabla_i$ acting on $\Psi(\boldsymbol{R}_e)$ is defined as
\begin{eqnarray}
       \frac{\nabla_i}{2}\Psi(\boldsymbol{R}_e) &= \frac{1}{2}\[ \Psi(\boldsymbol{R}_e + \hat{i} ) - \Psi(\boldsymbol{R}_e - \hat{i} ) \]
\end{eqnarray}
This describes two flavors of free, massless Dirac fermions in the low-energy region, since the second term in Eq.~\eqref{Hamiltonian-matrix-2-flavor} vanishes in the continuum limit.
A careful analysis in momentum space shows that the conserved non-singlet charges act as axial charges for the low-energy modes.
Using the expression after Stern transformation, we find 
 \begin{align}
    \label{eq:SU2A-transform}
    \lim_{N \to \infty}[Q_{\hat{x}_i}, \widetilde{\Psi}(\boldsymbol{k})] = (\gamma_5 \otimes \sigma_i) \, \widetilde{\Psi}(\boldsymbol{k}) \, ,
\end{align}
where $\gamma_5$ and $\sigma_i$ act on the spin indices and on the flavor indices of the fermion, respectively.
Each conserved charge $Q_{\hat{x}_i}$ can be regarded as the generator of a $\mathrm{U}(1)_{F_i}$ subgroup of $\mathrm{SU}(2)_L \times \mathrm{SU}(2)_R \times \mathrm{U}(1)_A$.

\section{Continuum Limit and Anomaly Discussion}
\label{sec:anomaly}

Although the lattice algebra of conserved charges appears to allow for
ultraviolet-scale anomalies, we do not expect any anomaly associated with
the flavor non-singlet chiral symmetry in the continuum limit.
This expectation is supported by perturbative analyses of the
Ward-Takahashi identities.

Requiring invariance under the complete set of lattice symmetries
constrains the theory to remain gapless, in accordance with continuum
considerations.
Hence, the anomaly-like features observed on the lattice should not be
interpreted as genuine infrared anomalies.


\subsection{Ons1 and anomaly}

It is well known that, in the continuum theory, there is no mixed anomaly
between $\mathrm{U}(1)_V$ and $\mathrm{U}(1)_{F_i}$.
Therefore, one anticipates the existence of a Hamiltonian deformation that
preserves both $Q_0$ and $Q_{\hat{x}_i}$ and lifts the fermionic zero modes.

For simplicity, and without loss of generality, we restrict ourselves to
Hamiltonians that preserve only $Q_0$ and $Q_{\hat{z}}$.
In other words, we focus on the potential mixed anomaly between
$\mathrm{U}(1)_V$ and $\mathrm{U}(1)_{F_3}$.
The symmetry $\mathrm{U}(1)_{F_3}$ generated by $Q_{\hat{z}}$ may be viewed
as a non-singlet chiral $\mathrm{U}(1)$ symmetry.

As expected, an explicit example of such a Hamiltonian deformation is
given by~\footnote{
We thank Elijah Lew-Smith and Shu-Heng Shao for bringing this to our attention.}
\begin{align}
\label{eq:Hamiltonian-def-Qz}
    M = \frac{\i}{2} m \sum_{<\boldsymbol{r},P[\boldsymbol{r}]>}
        \left(
        a_{\boldsymbol{r}} a_{P[\boldsymbol{r}]}
        + b_{\boldsymbol{r}} b_{P[\boldsymbol{r}]}
        \right) 
      = - m \sum_{\boldsymbol{R}_e}
        \Psi^{\dagger}(\boldsymbol{R}_e)
        \left( \gamma_0 \gamma_5 \otimes \i \sigma_2^{T} \right)
        \Psi(\boldsymbol{R}_e) .
\end{align}
Here,
$    <\boldsymbol{r}, P[\boldsymbol{r}]>
    = <(n_x,2n_y,n_z), \, (n_x,2n_y+1,n_z)> $.

One can verify explicitly on the lattice that~\footnote{
Other Hamiltonian deformations invariant under $Q_0$ and $Q_{\hat{z}}$
are also possible.
For instance, one may adopt the hopping pattern in
Eq.~\eqref{eq:Hamiltonian-def-Qz} as
\begin{align}
    <\boldsymbol{r}, P[\boldsymbol{r}]>
    = <(2n_x,n_y,n_z), \, (2n_x+1,n_y+1,n_z)> .
\end{align}
Such deformations break parity symmetry.
}
\begin{align}
     [ Q_0, M ] = [ Q_{\hat{z}}, M ] = 0 .
\end{align}

These results demonstrate that there is no mixed anomaly between
$\mathrm{U}(1)_V$ and $\mathrm{U}(1)_{F_3}$, even at the lattice level.
Furthermore, by evaluating the Ward--Takahashi identity associated with
the $\mathrm{U}(1)_{F_3}$ symmetry in the presence of a background gauge
field for the vector $\mathrm{U}(1)_V$ symmetry, one finds that lattice
symmetry-breaking effects do not induce relevant or marginal operators
that would signal an anomaly.
This observation is consistent with the well-established fact that
non-singlet chiral symmetries are anomaly-free in quantum field theory.

\section{Summary and Outlook}
\label{sec:summary}

We have studied conserved charges of the staggered fermion Hamiltonian in
$3+1$ dimensions.
Our main results are as follows:

\begin{itemize}
\item We constructed three independent conserved non-singlet charges.
\item These charges generate axial $SU(2)_L \times SU(2)_R$ transformations
in the continuum limit.
\item Lattice-specific non-commutativity of the charge algebra was
clarified and shown not to imply an infrared anomaly.
\end{itemize}

Future work includes coupling the system to dynamical gauge fields and
applying the present framework to numerical lattice simulations.









\acknowledgments
T.Y. sincerely thanks Soichiro Shimamori and Hiroki Wada for the many fruitful discussions, which were instrumental in shaping the direction and progress of this work.
The work of T.O. is supported in part by JSPS KAKENHI Grant Number 23K03387.
The work of T.Y. is supported in part by JST SPRING, Grant Number JP- MJSP2138.

\appendix

\bibliographystyle{unsrt}
\bibliography{reference}

\end{document}